\documentclass[aps,prb,twocolumn,groupedaddress,showpacs,floatfix]{revtex4}

\usepackage{bm}
\usepackage{amssymb}
\usepackage{amsmath2000}
\usepackage{graphicx}

\newcommand{\be}{\begin{equation}}
\newcommand{\fe}{\end{equation}}
\newcommand{\bmult}{\begin{multline}}
\newcommand{\fmult}{\end{multline}}

\newcommand{\rb}{\right}
\newcommand{\lb}{\left}

\newcommand{\Ref}[1]{Ref.\ \onlinecite{#1}}
\newcommand{\Refs}[1]{Refs.\ \onlinecite{#1}}

\newcommand{\bfig}{\begin{figure}}
\newcommand{\efig}{\end{figure}}  
\newcommand{\fig}[2]{\scalebox{#1}{\includegraphics*{#2}}}

\newcommand{\etal}{\textit{et al.}}
\newcommand{\hTc}{high-$\mathrm{T_c}$}

\newcommand{\bk}{{\mathbf k}}
\newcommand{\bK}{{\mathbf K}}
\newcommand{\br}{{\mathbf r}}

\newcommand{\ad}{\alpha_D}
\newcommand{\vd}{v_{\Delta}}
\newcommand{\dgr}{^{\dag}}
\newcommand{\ybcod}{$\mathrm{YBa_2Cu_3O_{7-\delta}}$}

\newcommand{\tbco}{$\mathrm{Tl_2Ba_2CuO_{6+\delta}}$}
\newcommand{\rate}{$T_1^{-1}$}
\newcommand{\TT}{$(T_1T)^{-1}$}

\begin{document}

\title{Dirac quasiparticles and spin-lattice relaxation in the mixed state}

\author{Daniel Knapp}
\email[Corresponding author: ]{dknapp@physics.mcmaster.ca}
\author{Catherine Kallin}
\author{A.\ John Berlinsky}
\affiliation{Department of Physics and Astronomy, McMaster University, Hamilton, Ontario, Canada L8S 4M1}

\author{Rachel Wortis}
\affiliation{Physics Department, Trent University, Peterborough, Ontario, Canada K9J 7B8}

\date{\today}

\begin{abstract}
We present the results of quantum-mechanical calculations, using the singular gauge transformation of Franz and Te\u{s}anovi\'{c}, of the rate of planar Cu spin-lattice relaxation due to electron spin-flip scattering in the mixed state of {\hTc} cuprate superconductors.  The results show a non-monotonic temperature and frequency dependence that differs markedly from semiclassical Doppler-shifted results and challenges the assertion that recent experimental observations of the rate of planar Cu and O spin-lattice relaxation in the mixed state of {\ybcod} point to antiferromagnetic spin fluctuations as a better candidate for the elementary excitations of the superconducting state.   
\end{abstract}

 %pacs numbers identify the categories of s/c state & response of superconductors to electromagnetic fields (eg. NMR) respectively

\maketitle

The nature of the low-lying excitations in the mixed state of $d$-wave superconductors remains an open question despite extensive research over the past several years.  Recent experimental\cite{curro_NMR, mitrovic_NMR_Nature, mitrovic_NMR_AF, kakuyanagi_NMR, kakuyanagi_NMR_Tl} and theoretical studies\cite{wortis_NMR, morr_wortis, morr_NMR, takigawa_NMR} have shown that the frequency dependence of the planar Cu and O spin-lattice relaxation rate (\rate) allows NMR to be used as a local probe of these low-lying excitations, offering a powerful test of theoretical treatments of the electronic structure in the mixed state.  

A generic feature of the NMR experiments is that {\rate} shows a strong frequency dependence, increasing sharply with increasing resonance frequency, i.e.\ with decreasing distance from the vortex cores.  This frequency dependence has been qualitatively reproduced within a semiclassical model of Doppler-shifted Dirac quasiparticles by Wortis {\etal}\cite{wortis_NMR} (WBK).  However, experiments on the O spins in \ybcod\cite{mitrovic_NMR_Nature, mitrovic_NMR_AF, kakuyanagi_NMR} and on the Tl spins in {\tbco}\cite{kakuyanagi_NMR_Tl} have shown that the relaxation rate has a non-monotonic dependence on resonance frequency, first decreasing as one moves away from the vortex cores and then increasing as one approaches the minimum resonance frequency, precisely where the Doppler shift goes to zero.  This is in clear disagreement with the semiclassical picture where the Doppler shift of the linear dispersion of nodal quasiparticles is what gives rise to a non-zero local density of states at low temperatures. An observed non-monotonic dependence of {\TT} on temperature\cite{curro_NMR, mitrovic_NMR_AF, kakuyanagi_NMR_Tl} is also in disagreement with the semiclassical model\cite{wortis_NMR} which predicts that {\TT} will always increase with temperature.   
Different theoretical predictions were made by Morr and Wortis\cite{morr_wortis} and Morr\cite{morr_NMR} who calculated  {\rate} within a spin-fermion model in which the damping of antiferromagnetic spin fluctuations is determined by their coupling to planar quasiparticles.  At low temperatures they found that {\rate} increases monotonically with resonance frequency, but they also found that as the temperature is increased, {\TT} actually decreases at the high-frequency end of the curve and increases at the low-frequency end such that a minimum develops at a temperature-dependent crossover point.  Morr and Wortis noted that this temperature and frequency dependence is different from that calculated within the Dirac quasiparticle model and proposed that these differences would allow NMR experiments to determine whether it is Dirac quasiparticles or strong spin fluctuations that govern the low-temperature behaviour in the superconducting state of the {\hTc} cuprates.  In a later paper, where Morr repeated the calculations of \Ref{morr_wortis} for the planar O spins, Morr further argued, based on comparisons with the experimental results of Curro {\etal}\cite{curro_NMR}, that the antiferromagnetic spin-fluctuation mechanism is the dominant contributor to the relaxation rate for both the O and Cu spins.   

In this paper we show that a quantum-mechanical treatment of Dirac quasiparticles leads to results that are qualitatively different from those of the semiclassical treatment of \Ref{wortis_NMR}.  In particular, we find that {\rate} has a non-monotonic dependence on temperature and on resonance frequency.  These results complicate the experimental distinction between Dirac quasiparticles and antiferromagnetic spin fluctuations, and suggest that the experimentally observed relaxation rate outside of the vortex core can be largely explained (depending on the orientation of the vortex lattice) by the influence of the periodic vortex lattice on the Dirac quasiparticles.

In what follows we confine ourselves to a single Cu-O plane, and assume that the electronic properties are entirely two-dimensional.  Following WBK, we consider the ($-\tfrac{1}{2}\leftrightarrow-\tfrac{3}{2}$) transition of the in-plane Cu atoms and write the relaxation rate as
%equation
\be \label{eq_t1_1}
\begin{split}
T_1^{-1}&(\br)=\frac{1}{3} \frac{2\pi}{\hbar} (\gamma_e\gamma_n\hbar^2) 
|\langle-\tfrac{1}{2}|I_{+}(\br)|-\tfrac{3}{2}\rangle|^2 \\
\times&
\biggl\langle 
\Bigl|
\langle f|A_{\perp}S_{-}(\br)+\sum_{\nu}B_{\perp}S_{-}(\br+\bm{\delta}_{\nu})|i\rangle
\Bigr|^2 
\delta(E_i-E_f)
\biggr\rangle,
\end{split}
\fe
where $i$ and $f$ are the initial and final many body states of the electronic system, $\br$ is the position of a copper atom and $\bm{\delta}_{\nu}$ are the displacements to the four nearest-neighbor copper atoms. $\langle \ \rangle$ denotes a thermal average, $\gamma_e\gamma_n\hbar^2B_{\perp}\approx3.06\times10^{-19}$ erg, and $A_{\perp}/B_{\perp}\approx 0.8$. We neglect the nuclear Zeeman splitting in the $\delta$ function, because it is so much smaller than that of the electron.  The nuclear matrix element, with $I_{+}(\br)$, simply gives a factor of 3.  The spin operators
%equation
\be \label{eq_spin_op}
S_{-}(\br)=\Psi_{\downarrow}\dgr(\br)\Psi_{\uparrow}(\br)
\fe
are expanded in terms of Bogoliubov quasiparticle operators
%equation
\be
\Psi_{\uparrow}(\br)=
\frac{1}{\sqrt{N_{\perp}}}
\sum_{\mu n \bk}[\gamma_{\mu n \bk\uparrow}u_{\mu n \bk\uparrow}(\br)-\gamma_{\mu n \bk\downarrow}\dgr v_{\mu n \bk\downarrow}^{*}(\br)],
\label{eq_quasiparticles}
\fe
where the sum is to be taken over the node $\mu$, the band index $n$ and the wavevector $\bk$, and $N_{\perp}$ is the number of planar Cu sites.    

For a $d$-wave superconductor in the mixed state Franz and Te\u{s}anovi\'c\cite{franz_tesanovic} have used a bi-partite singular gauge transformation (see Fig.\ \ref{fig_square_lattice})
%figure
\bfig
\fig{0.55}{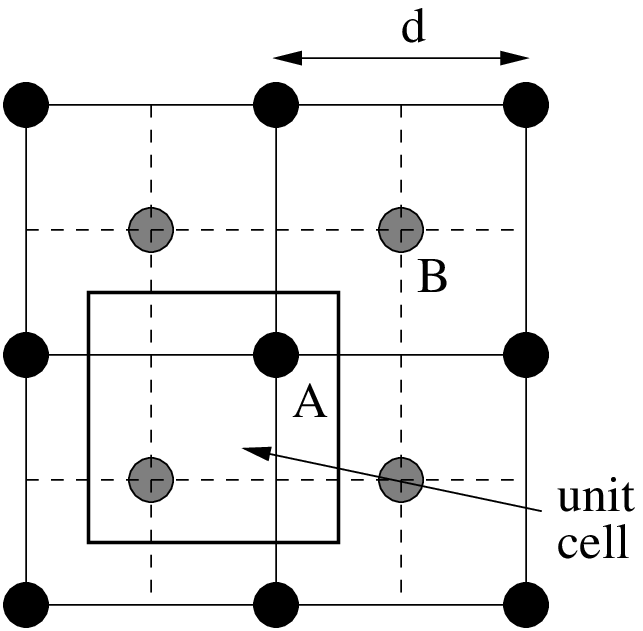}
\caption{\label{fig_square_lattice} The square vortex lattice used in this paper, divided into two sublattices.  Note that the $x$- and $y$-axes are chosen to lie along the node directions.}
\efig
 to show that, close to the nodes, the quasiparticle wavefunctions are best described by Bloch waves propagating through a periodic potential created by the combined action of the applied magnetic field and the periodic structure of supercurrents winding around a lattice of vortices.  Note that we assume the London model, neglecting variation in the gap magnitude and the local magnetic field, and treat the vortex cores as point objects.  For large values of the Dirac cone anisotropy ($\ad=v_F/\vd$, where $v_F$ is the Fermi velocity at the node and $\vd$ is the slope of the gap at the node) Mel'nikov\cite{melnikov_1D} showed that the quasiparticle wavefunctions are confined by the vortex lattice to lie along the nodal directions.  Knapp {\etal}\cite{knapp_Dirac} further showed that Mel'nikov's 1D approximation could be used to calculate an energy spectrum for the quasiparticles in excellent qualitative agreement with that calculated by Franz and Te\u{s}anovi\'c\cite{franz_tesanovic}.  Knapp {\etal} also showed that the approximation could be improved by adding small numbers of plane waves along the nodal direction in a ``quasi-1D'' approximation.  We use this approximation here to expand the quasiparticle wavefunctions around each node as a sum over plane waves.  For example, at the node $(k_F,0)$ (in this paper we set the nodes to lie along either the $x$- or $y$-axis) we write
%equation
\be \label{eq_planewave}
u_{n k_x \alpha}^{(k_F,0)}(\br)=
e^{i p_F x} 
\sum_{K_y}^{K_{\smash[b]{y}}^c}\sum_{K_x}^{K_x^c}
U_{n k_x-\bK \alpha}^{(k_F,0)}
e^{i(k_x-\bK) \cdot \br}
\fe
where $p_{F}$ is the Fermi momentum at the node, $\alpha$ is the spin and
$\bK$ are the reciprocal lattice vectors of the vortex lattice.  $K_y^c$ is the cutoff wave vector along the $y$ (transverse) direction and $K_x^c$ is the cutoff wave vector along the $x$ (node) direction and is taken to be smaller than $K_y^c$ in the quasi-1D approximation.  Note that for computational convenience we neglect the $k_y$ dependence, a good approximation at low energies.  

If we substitute the plane wave expansion for the quasiparticle wavefunctions into Eq.\ \eqref{eq_quasiparticles}, and perform the thermal average we find that 
%single column
\begin{widetext}
%equation
\be \label{eq_t1_r}
\begin{split}
T_1^{-1}(\br)=& \frac{2 \pi}{\hbar}
\left( \gamma_e \gamma_n \hbar^2 \right)^2 \frac{1}{N_{\perp}^2}  
\sum_{\mu n \bk}\sum_{\mu' n' \bk'}
f(\epsilon_{\mu n \bk \uparrow})
[1-f(\epsilon_{\mu' n' \bk' \downarrow})] 
\left[ A_{\perp} + 4 B_{\perp} \cos \lb( (\mu -\mu')\frac{\pi}{2} \rb) \right]^2 
\\
&\times
\delta(\epsilon_{\mu n \bk \uparrow}-\epsilon_{\mu' n' \bk' \downarrow})
\left| 
\sum_{\bK \bK'}
\left( U^{*}_{\mu n \bk-\bK \uparrow}
       U_{\mu' n' \bk'-\bK' \downarrow}
       +
       V^*_{\mu n \bk-\bK \uparrow}
       V_{\mu' n' \bk'-\bK' \downarrow}
\right)
e^{+i(\bK-\bK') \cdot \br} \right|^2.
\end{split}
\fe
\end{widetext}
Note that we have assumed that we are always close enough to the nodes that
%equation
\begin{multline}
\cos \left(a[(p_{\mu x}-p_{\mu'x})+(k_x-K_x)-(k_x'-K_x')]\right)\\
\sim
\cos \left(a[p_{\mu x}-p_{\mu'x}]\right),
\end{multline}
where $a\simeq 3.855$ {\AA} is the distance between adjacent copper atoms.  In   the calculations presented here, we have used the numbers in Chiao {\etal}\cite{chiao} for YBCO:  $v_F \simeq 2.5 \times 10^7$ cm/s, $\ad=14$. In the plane wave expansion of Eq.\ \eqref{eq_planewave} we have used 45 reciprocal lattice vectors along the transverse direction and 11 along the node direction.  

The calculated spatial structure of {\rate} is shown in Fig.\ \ref{fig_t1_r}.  
As in the semiclassical calculation\cite{wortis_NMR}, the rate is highest near the vortex cores and falls with increasing distance from the core.   What is not seen in the semiclassical calculation is the complicated four-fold spatial structure of {\rate}.  This structure is entirely a quantum effect which is due to the extension, at low energies, of the quasiparticle wavefunctions along the node directions and by the confinement of the quasiparticle wavefunctions by the periodic potential of the vortex lattice.  The degree of spatial anisotropy increases with the size of the Dirac cone anisotropy, $\ad$.    Note that this structure is not due to variation in the gap magnitude at the vortex core, which we neglect here.  The extension of quasiparticle states along the node directions is observed in other theoretical treatments of quasiparticles in the vortex lattice\cite{ichioka_vortex_lattice, melnikov_STM, vafek}, and the analogue of these states for the case of a single vortex is discussed in \Refs{franz_single_vortex, melnikov_a_bohm, han_lee}.
%figure
\bfig
\fig{0.55}{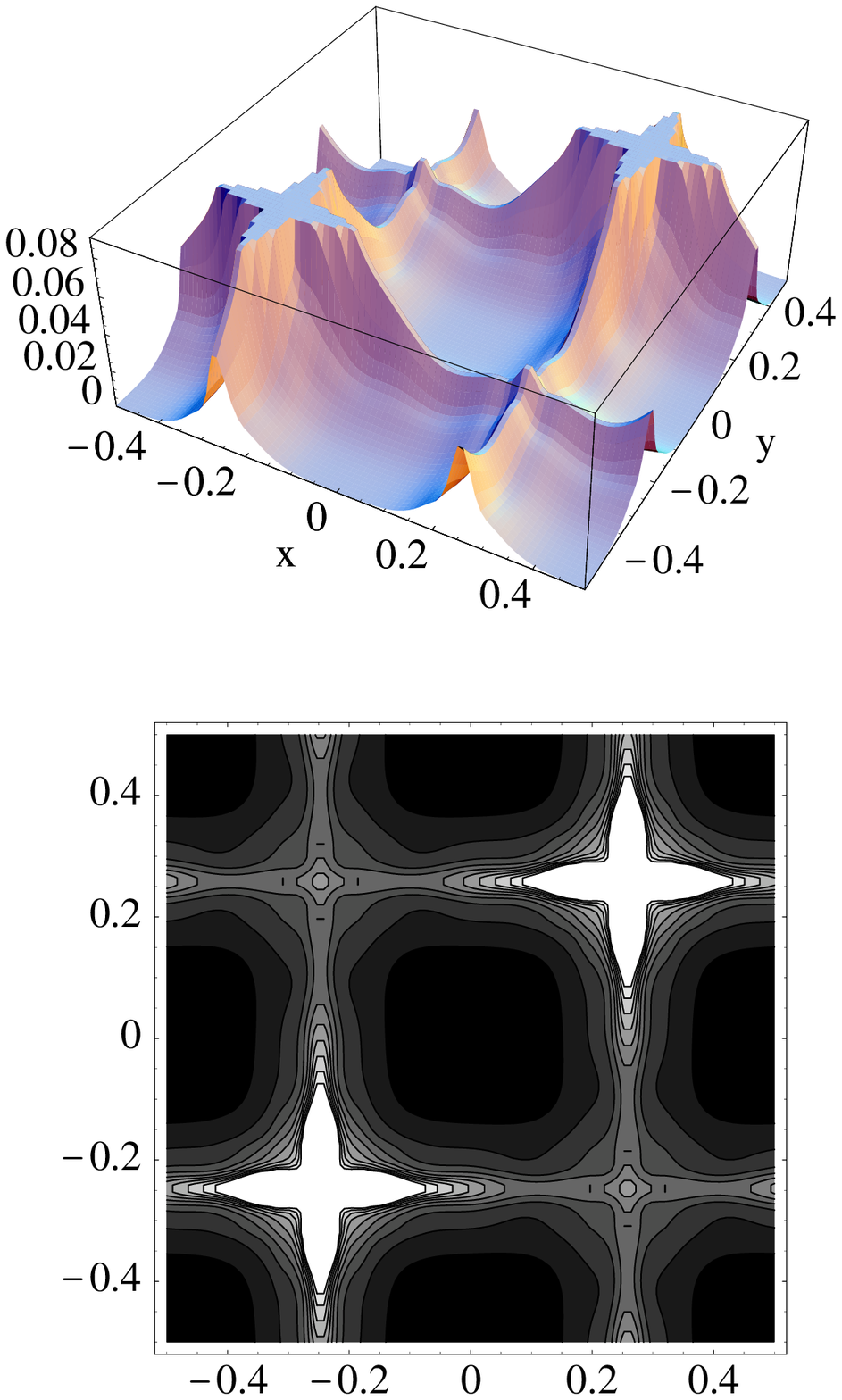}
\caption{\label{fig_t1_r}$(T_1(\br)T)^{-1}$ in a two-vortex unit cell at 9 K in an applied field of 16 T.  The peaks at the vortex positions have been cut off in the top panel to better show the spatial structure.}
\efig

The position-dependent rate can be used to generate a frequency-dependent rate (shown in Fig.\ \ref{fig_t1_f}) 
%figure
\bfig
\fig{0.33}{figure3.eps}
\caption{\label{fig_t1_f}{\TT} as a function of local magnetic field, $B$ (or resonance frequency, $\omega\propto B$) for $\text{H}=\text{8 T}$ ($\diamond$), $\text{H}=\text{16 T}$ ($\circ$) and $\text{T}=\text{9 K}$.  Note the upturn in the rate at the minimum in the local field (frequency), and the large spread in {\rate}.}
\efig
by correlating the position in the vortex lattice with the local magnetic field, as calculated using the London model:
%equation
\be
B_z(\br)=H\sum_{\bK}\frac{e^{i\bK \cdot \br}}{1+\lambda^2 K^2}e^{-\xi^2 K^2/2},
\fe
where we have used $\lambda\approx 1600$ {\AA} and $\xi\approx 16$ {\AA}.  As noted already, the rate increases with increasing frequency, however there is also an upturn at the minimum frequency.  This upturn is due to the intersection of the extended quasiparticle wavefunctions with the local field minimum (see Figs.\ \ref{fig_t1_r}, \ref{fig_lattice_structure}).  
%figure
\bfig
\fig{0.5}{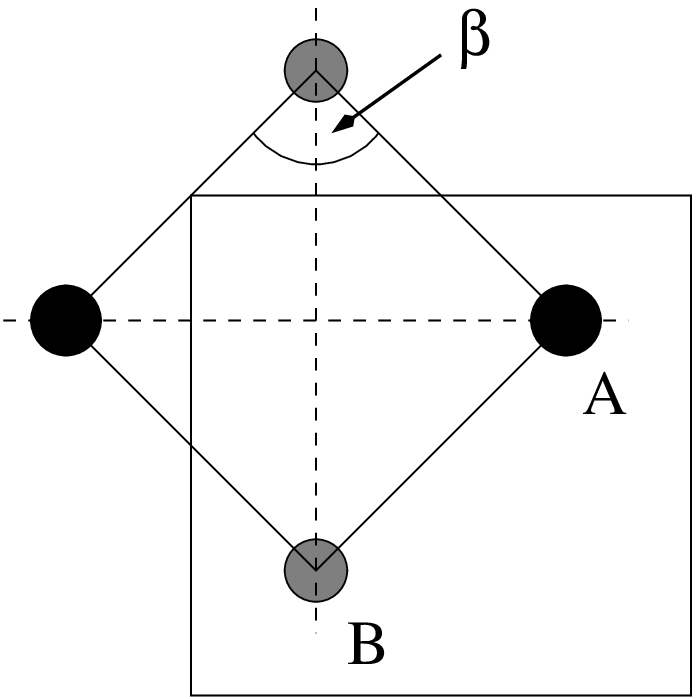}
\caption{\label{fig_lattice_structure}  In order to have an upturn in {\rate} at low frequencies lines drawn from the vortices along the node directions must cross the local field minimum.}
\efig
It is in this sense that the low-frequency dependence is sensitive to the orientation of the vortex lattice.  In the case of a square vortex lattice, this effect will be seen if the lattice is oriented along the Cu-O bonds.

For applied magnetic fields below $\text{6 T}$ the vortex lattice is believed to be triangular or oblique\cite{yethiraj_SANS_twinned, keimer_SANS_twinned, johnson_SANS_untwinned, maggio_STM_lattice, maggio_STM_twins}, in which this effect might be seen for certain orientations.  There is also, however, strong experimental\cite{gilardi_SANS_LSCO} and theoretical\cite{berlinsky_GL,franz_London, won_maki, ichioka_Eilenberger, mandal_ramakrishnan} evidence for a transition from a triangular to a square vortex lattice with increasing magnetic field.  While the theoretical results suggest that the square lattice should be oriented along the node directions, the recent small-angle neutron scattering results for LSCO\cite{gilardi_SANS_LSCO} see very clear evidence for a field-based transition between 0.5 and 0.8 T to a square lattice that is oriented along the Cu-O bonds. Due to the smaller gap in LSCO, Gilardi {\etal} suggest that a similar transition might be seen in YBCO for fields a factor of 10 higher, on the order of 5 to 8 T.  While the exact structure and orientation of the vortex lattice remain controversial, we would argue that there is good evidence for the type of lattice in which Dirac quasiparticles would contribute to a low-frequency upturn in {\rate}.  An excellent test of this theory would be a frequency-dependent measurement of {\rate} in a material known to have a vortex lattice which does not have the orientation shown in Fig. \ref{fig_lattice_structure}.  

The temperature dependence of {\TT} at three different frequency positions in the unit cell is shown in Fig.\ \ref{fig_t1_t}.  
%figure
\bfig
\fig{0.33}{figure5.eps}
\caption{\label{fig_t1_t} {\TT} as a function of temperature at 16 T, averaged over a narrow frequency region near the core (solid line), at the minimum frequency (dashed line) and at the frequency where the rate goes to a minimum (dotted line).  The inset shows a close-up of the slower rates.}
\efig
At all positions, the relaxation rate starts from zero at $T=0$ and quickly rises to a peak near $T=2\ \text{K}$.  Away from the vortex, {\TT} then decreases until $T\sim 10-20\text{ K}$, where it starts to increase again.  This is in contrast with the spin-fluctuation results\cite{morr_wortis,morr_NMR} where {\TT} was found to increase with temperature at the low-frequency end and decrease at the high-frequency end of the spectrum.  The temperature dependence due to Dirac quasiparticles can be understood from the quasiparticle density of states.  Following from the work of Franz {\etal}\cite{franz_tesanovic}, Marinelli {\etal}{\cite{marinelli_Dirac} showed that the density of states starts from zero at zero energy, but increases very rapidly due to the existence of energy bands which are bent back toward the Fermi surface by the periodic potential of the vortex lattice.  Low energy structure in the density of states created by the bending of the lowest energy bands was further shown by Knapp {\etal}\cite{knapp_Dirac} to give a specific heat which exhibits similar behaviour to the temperature dependence of {\TT} shown here.  We note that this temperature dependence is similar to that seen experimentally for the O\cite{curro_NMR, mitrovic_NMR_AF} and Tl\cite{kakuyanagi_NMR_Tl} spin lattices and that, unlike the low frequency upturn in {\rate}, the temperature dependence is not dependent on the orientation of the vortex lattice. 
\vspace{0.1in} %had some trouble with figure placement               

From Figures \ref{fig_t1_f} and \ref{fig_t1_h} one can see that increasing the applied magnetic field has two effects: (i) at larger fields, the local magnetic field becomes more uniform and the frequency range narrows, but the overall shape of {\rate} stays the same (see Fig.\ \ref{fig_t1_f}); (ii) the low-lying density of states scales with $\sqrt{H}$ due to the existence of near-nodes in the quasiparticle energy spectrum\cite{franz_tesanovic, marinelli_Dirac, knapp_Dirac}, so that $\text{\TT}\sim H$ (see Fig.\ \ref{fig_t1_h}) in agreement with the experimental findings of Zheng {\etal}\cite{zheng_NMR} for $^{63}\mathrm{Cu}$ NMR in $\mathrm{TlSr_2CaCu_2O_{6.8}}$. 
%figure
\bfig
%\vspace{0.1in}
\fig{0.33}{figure6.eps}
\caption{\label{fig_t1_h} {\TT} averaged over the vortex unit cell as a function of applied magnetic field for $T=\text{1 K}$ ($\diamond$), 9 K ($\bigtriangleup$), and 19 K ($\circ$).  The structure near 15 T is due to shifting of the lowest quasiparticle energy bands which is caused by increasing the magnetic field.}
\efig

In conclusion, we have shown that the quantum-mechanical treatment of Dirac quasiparticles is able to qualitatively account for non-monotonic structure in the frequency and temperature dependence of {\TT} in the mixed state of {\hTc} cuprate superconductors.  The quantum-mechanical results differ markedly from the semiclassical results of \Ref{wortis_NMR} and show some similarities with the spin-fluctuation picture of \Refs{morr_wortis, morr_NMR}.  These similarities complicate the experimental distinction between Dirac quasiparticles and spin fluctuations and call into question the recent assertion\cite{morr_NMR} that antiferromagnetic spin fluctuations are the dominant mechanism for spin-lattice relaxation in the mixed state.  

\begin{acknowledgments}
D.\ K.\ would like to thank David Cooke for his timely programming advice.   We acknowledge the use of SHARCNET facilities at McMaster where the computing for this project was carried out (\url{http://sharcnet.mcmaster.ca}).   This research was supported by the Natural Sciences and Engineering Research Council (Canada) and by the Canadian Institute for Advanced Research.
\end{acknowledgments}

\end{document}